\documentclass{ws-ijmpe}

\begin{document}

\markboth{Authors' Names}{Instructions for  
Typing Manuscripts (Paper's Title)}

\catchline{}{}{}{}{}

\title{STAGGERING OF THE B(M1) VALUE AS A FINGERPRINT OF SPECIFIC CHIRAL BANDS STRUCTURE}

\author{\footnotesize E. GRODNER}

\address{Faculty of Physics, University of Warsaw, ul. Ho\.za 69, Warszawa 00-681, Poland
ernest.grodner@fuw.edu.pl}

\maketitle

\begin{history}
\received{(received date)}
\revised{(revised date)}
\end{history}

\begin{abstract}
Nuclear chirality has been intensively studdied for the last several years 
in the context of experimental as well as theoretical approach. 
Characteristic gamma selection rules have been predicted for the strong
chiral symmetry breaking limit that has been observed in Cs isotopes. The presented analysis
shows that the gamma selection rules cannot be attributed only to chiral symmetry breaking.
The selection rules relate to structural composition of the chiral rotational bands, i.e. to odd particle configuration and the deformation of the core.
\end{abstract}

\section{Introduction}
Chirality in nuclear physics relates to features of nuclear hamiltonian 
in the context of time reversal symmetry. The phenomenon of explicit time-reversal symmetry breaking
comes up through already well known Cabibbo-Kobayashi-Maskawa 
matrix of the standard model. The CKM matrix, introduced in order to diagonalize 
the mass terms given by Higgs mechanism with respect to quark flavor, contains T violating terms being 
associated with the heavy quark sector. Therefore no explicit T violation is expected to be present 
in low energy nuclear interactions giving T-symmetric nuclear hamiltonian.
What remains is the mechanism of spontaneous time reversal symmetry breaking and possible  
occurrence of spontaneous breakdown of the chiral symmetry.
This possibility has been studied for several years through the measurements of the energy levels and 
gamma transition probabilities \cite{grodner_06},\cite{tonev_06}. The latter ones have been intensively studied 
in the mass regions of A=130 and A=100, while the specific gamma selection rules, i.e. staggering of 
the B(M1) values along the rotational bands have been observed only in Cs isotopes \cite{grodner_06}. 
Absence of the B(M1) staggering in other nuclei is intriguing and needs clarification what is the main 
goal of the present paper.

\section{Eigenstates of the nuclear hamiltonian and phase convention}
Although spontaneous time-reversal symmetry breaking is responsible for the phenomenon of chirality, 
there are no eigenstates of T operator and therefore the symmetry operator T alone is not used in the 
quantum-mechanical description of this problem.
However, one can introduce another symmetry operator $R_y^T$ being a combination 
of time-reversal and $\pi$-rotation, that allows proper description of T-symmetry breaking 
and simultaneously possesses its own eigenstates. The $R_y^T$ is called T-signature \cite{droste} 
or chiral symmetry \cite{frauendorf} operator.

As stated above, nuclear hamiltonian H is expected to be T symmetric. Additionally,  
H is also rotationally invariant in the absence of external fields. As a result, the eigenstates of H are 
assigned to spin and spin projection quantum numbers
$|IM\rangle$, and the following commutation relation is valid
\begin{equation}
\label{commutation_1}
\left[R_y^T, H\right]=0.
\end{equation}
The commutation relation (\ref{commutation_1}) shows that if $|IM \rangle$ is 
an eigenstate of H with the eigenvalue $E$, then $R_y^T|IM\rangle$ must 
also be the eigenstate of H with the same eigenvalue
\begin{equation}
\label{eigenstates}
\langle IM|H|IM \rangle = \left( \langle IM|R_y^{T+}\right)HR_y^T|IM\rangle.
\end{equation}
The $R_y^T$ operator is anti-unitary the action of which makes complex conjugation of all complex numbers. 
Therefore, it does matter which part of the eqn. (\ref{eigenstates}) 
the operator acts on, and the appearance of the round bracket in the eqn. (\ref{eigenstates}) 
is necessary.
The round bracket shows that $R_y^{T+}$ acts only on the $\langle IM|$. Skipping 
this bracket equals to the complex conjugation of the whole matrix element
\begin{equation}
\langle IM|H|IM \rangle = \langle IM|R_yT^+HR_yT|IM\rangle^{\star}.
\end{equation}
Spin and its projection are not changed by acting of $R_y^T$ operator \cite{bohr-mottelson} ,i.e. 
\begin{equation}
R_y^T |IM\rangle = c|IM\rangle
\end{equation}
where $c$  is a complex number. The above equation shows that eigenstates of 
the hamiltonian are the eigenstates of the symmetry operator $R_y^T$ as well 
and, therefore, they preserve the chiral symmetry.
Owing to the fact that eigenvalues of the hamiltonian are real values, one can show 
that $c$ must be phase only ($|c|^2=1$) 
\begin{eqnarray}
\langle IM|H|IM \rangle &=& \\
&=&\langle IM|R_y^{T+}HR_y^T|IM\rangle^{\star}\\
&=& cc^{\star}   \langle IM|H|IM \rangle^{\star}\\
&=& |c|^2 \langle IM|H|IM\rangle.
\end{eqnarray}
The above analysis relates to diagonal matrix elements of the hamiltonian but it can be generalized.
Particularly, it is possible to make any matrix element of any operator $\mathcal{M}$ 
commuting with $R_y^T$ to be real by choosing the phase $c=1$
\begin{equation}
\label{creversed}
\langle I_2M_2 |\mathcal{M}| I_1M_1 \rangle = c_2c_1^{\star} \langle I_2M_2 |\mathcal{M}| I_1M_1 \rangle^{\star} = \langle I_2M_2 |\mathcal{M}| I_1M_1 \rangle^{\star}.
\end{equation} 
The choice of this special phase $c=1$ introduces a phase convention \cite{bohr-mottelson}, 
and this convention has to be kept for the further analysis.

\section{Symmetry breaking states}
In the most simple case of chirality, the nuclei are described as three-body objects being composed of two odd nucleons and an even-even core.
Therefore presence of three angular momenta vectors forming a system with definite handedness is expected in tandem with specific conditions 
where one of the 
odd nucleons possesses particle-like character, the other one hole-like character and a core is triaxially deformed.
For that reason, apart of the eigenstate basis $|IM\rangle$ one can introduce equivalent basis set in which the handedness of the nucleus is specified.
Although the handedness, being dependent on the angular momenta vectors orientation, is a continuous parameter, it can be classified in two values $L$ and $R$
corresponding to the left- or right-handedness of the system. 
The elements of the new basis set have several interesting features. 
In contrast to the matrix elements calculated on the $|IM\rangle$ states being real, 
the matrix elements calculated on the $|L\rangle$, $|R\rangle$ are complex since $|L\rangle$ and $|R\rangle$ are not the eigenstates of the hamiltonian.
This means also that they relate to dynamical features of the nucleus.
The states with specified handedness break the chiral symmetry since they are not the eigenstates of the $R_y^T$ symmetry operator i.e.
\begin{eqnarray}
\label{chiral1}
R_yT |L\rangle &=& |R\rangle\\
\label{chiral2} 
R_yT |R\rangle &=& |L\rangle.
\end{eqnarray}  

\section{Symmetry restoration}
The $|L\rangle$ and $|R\rangle$ not being the eigenstates of the hamiltonian posses an essential dynamical feature, i.e. 
the system can tunnel between 
left- and right-handed configurations 
\begin{equation}
\Delta E=\langle L|H|R\rangle \neq 0.
\end{equation}
One can estimate the tunneling frequency as $f\approx 2.5\cdot 10^{19}$Hz by taking the ratio $f = \Delta E / h $, where $\Delta E=100$ keV is typical energy splitting observed in the chiral nuclei (see following sections).
Since the time of gamma quanta emission is larger than the tunneling period, it is not possible to observe such fast 
dynamical process experimentally through electromagnetic radiation. 
In this case the time scale of the measurement is much larger than the characteristic time of the tunneling process.
Following the indeterminacy principle, the states with definite energy and undefined time 
are observed experimentally, 
i.e. only the eigenstates of the hamiltonian $|IM\rangle$ are visible.
Therefore the $|L\rangle$ and $|R\rangle$ states have to be projected on $|IM\rangle$ basis elements in order to compare 
the theoretical predictions with the experimental observations. The projection procedure is also called symmetry restoration. In the case of the chiral symmetry
the rotational as well as the T-symmetry (chirality) has to be restored. 

Rotational symmetry can be restored in the spirit of Generator Coordinate Method method (see Sect. 11.4.6 in Ref. \cite{ring-schuck}) by using
projection operators $P_I^M$ 
\begin{eqnarray}
\label{rzutowanko1}
P_I^M|L,n\rangle&=&|L,I,M,n\rangle \\
\label{rzutowanko2}
P_I^M|R,n\rangle&=&|R,I,M,n\rangle
\end{eqnarray}
where the handedness is not affected by spin projection since there is no $R_y^T$ in the operators $P_I^M$.
Indices $n$ in Eqs. (\ref{rzutowanko1}) and (\ref{rzutowanko2}) stand for quantum numbers -- like spin of the core $j_c$ and 
of the odd nucleons $j_{\nu}$, $j_{\pi}$ etc. -- related to structural composition of the partner band levels. 
These quantum numbers will be omitted in order to examine the role of the handedness only.
The spin quantum number is not affected by action of $R_y^T$ operator  
since according to Eq.(1-39) of Ref.\cite{bohr-mottelson} it fulfills the following commutation relations
\begin{eqnarray}
\label{kolejnosc1}
\left[ R_y^T,I_z \right] &=&0 \\
\label{kolejnosc2}
\left[ R_y^T,I^2 \right] &=&0.
\end{eqnarray}
Therefore the restoration of the chiral symmetry (projection of chirality quantum number $+$ or $-$) have the following form
\begin{eqnarray}
|IM,+\rangle &=& \frac{1}{\sqrt{2}N_+}(|IM,L\rangle + |IM,R\rangle)\\
|IM,-\rangle &=& \frac{i}{\sqrt{2}N_-}(|IM,L\rangle - |IM,R\rangle)
\end{eqnarray}
$N_{\pm}$ being the normalization factors.
The $|IM,\pm\rangle$ states form chiral doubles that can be observed experimentally.
These doublets form two chiral partner bands.
Phases of $|IM,\pm\rangle$ states have been chosen so as to fulfill the phase convention given by Eq. (1-38) of 
Ref. \cite{bohr-mottelson}.
The formulae (\ref{kolejnosc1},\ref{kolejnosc2}) show that the order of the projections can be reversed and one can project on the chirality first and then on the spin quantum number.

\section{Electromagnetic transition matrix elements}
The probabilities of the electromagnetic transitions linking the excited states in the chiral bands relate to the matrix elements of the 
electromagnetic transition operator $\mathcal{M}(\sigma\lambda)$ calculated on the laboratory states $|IM,\pm \rangle$.
As it was shown in
Ref. \cite{grodner_07}, the $\mathcal{M}(\sigma\lambda)$ fulfills the following commutation
relation (see also Eqs. (1A-73) and (3C-10) in Ref. \cite{bohr-mottelson}) 
\begin{equation}
\label{emreversed}
\left[ R_yT, \mathcal{M}(\sigma\lambda)\right] = 0
\end{equation}
$\sigma\lambda$ being the M1, E2, M3, E4,... transition type.
This commutation relation, together with the introduced phase convention makes all matrix elements of the electromagnetic transitions between the 
$|I,M\pm\rangle$ states real values
and that for the 
inband electromagnetic transition $\langle I_2,\pm||\mathcal{M}||I_1,\pm\rangle$, i.e. without chirality change, 
will assume the following form                                                                       
\begin{eqnarray}
\label{inbandmt}                                                                                        
\frac{\langle L,I_2||\mathcal{M}||L,I_1\rangle + 
\langle R,I_2||\mathcal{M}||R,I_1\rangle}{2N_{I_2\pm}N_{I_1\pm}}\\
\nonumber                              
\pm\frac{ \langle L,I_2||\mathcal{M}||R,I_1\rangle +                                          
\langle R,I_2||\mathcal{M}||L,I_1\rangle}{2N_{I_2\pm}N_{I_1\pm}}
\end{eqnarray}
and, accordingly, for the $\langle I_2,\pm||\mathcal{M}||I_1,\mp\rangle$ interband transition that changes chirality  
\begin{eqnarray}
\label{interbandmt}                           
i \frac{\langle L,I_2||\mathcal{M}||L,I_1\rangle -         
\langle R,I_2||\mathcal{M}||R,I_1\rangle}{2N_{I_2\pm}N_{I_1\mp}} \\
\nonumber
\mp i\frac{ \langle L,I_2||\mathcal{M}||R,I_1\rangle -                                        
\langle R,I_2||\mathcal{M}||L,I_1\rangle}{2N_{I_2\pm}N_{I_1\mp}}.                              
\end{eqnarray}                                                                                                        
One has to remember that matrix elements of $|I,L\rangle$ and $|I,R\rangle$ states take in general complex values.
Relations (\ref{creversed}), (\ref{chiral1}) and (\ref{chiral2}), (\ref{emreversed}) lead to the following expressions
\begin{eqnarray}
\langle I_2,L||\mathcal{M}||I_1,L\rangle^{\star}&=&\langle I_2,R||\mathcal{M}||I_1,R\rangle\\
\langle I_2,L||\mathcal{M}||I_1,R\rangle^{\star}&=&\langle I_2,R||\mathcal{M}||I_1,L\rangle.
\end{eqnarray}
The above relations show that summation and subtraction of complex conjugate values occurs in the matrix 
elements given by Eqs.(\ref{inbandmt}) and (\ref{interbandmt}). Therefore, the inband matrix element,
$\langle I_2\pm|\mathcal{M}|I_1\pm\rangle$, takes the final form 
\begin{equation}
\label{innband}
\frac{{\mathrm Re}\langle L,I_2||\mathcal{M}||L,I_1\rangle
\pm {\mathrm Re}\langle L,I_2||\mathcal{M}||R,I_1\rangle}
{N_{I_2\pm}N_{I_1\pm}}
\end{equation}
and the interband matrix element, $\langle I_2\pm|\mathcal{M}|I_1\mp\rangle$, transforms to
\begin{equation}
\label{innterband}
\frac{{\mathrm Im}\langle L,I_2||\mathcal{M}||L,I_1\rangle
\mp {\mathrm Im}\langle L,I_2||\mathcal{M}||R,I_1\rangle}
{\mp N_{I_2\pm}N_{I_1\mp}}.
\end{equation}
No imaginary unit appeares in equations (\ref{innband}),(\ref{innterband}) which is due to the
the chosen phase convention.
These equations are valid when the chiral symmetry is close to the strong symmetry breaking limit.
When this limit is attained, the tunneling effect between the left- and right-handed
states disappears, which means that $\langle L,I|H|R,I\rangle=0$.
The left- and right-handed states are separated $\langle L,I|R,I\rangle=0$
which means that normalization parameters $N_{I\pm}$ equal to unity.
Gamma transitions of M1 and E2 type cannot change right-handed state to the left-handed one i.e. 
$\langle L,I_2||\mathcal{M}(\sigma\lambda)||R,I_1\rangle=0$ ($\sigma\lambda$= M1, E2) 
since larger angular momentum change of the nuclear state would be required. In this case 
Eqs. (\ref{innband}, \ref{innterband}) reduce to
\begin{eqnarray}
\label{inband}
\langle I_2,\pm||\mathcal{M}||I_1,\pm\rangle &=& {\mathrm Re}\langle L,I_2||\mathcal{M}||L,I_1\rangle\\
\label{interband}
\langle I_2,\pm||\mathcal{M}||I_1\mp\rangle &=& \mp {\mathrm Im}\langle L,I_2||\mathcal{M}||L,I_1\rangle
\end{eqnarray}
$\mathcal{M}$ being the M1 or E2 electromagnetic transition operator.
The above equations show that in the case of a strong chiral symmetry breaking limit, the B(M1) and B(E2) transition 
probabilities in both bands should be identical. 
Moreover, equations (\ref{inband}) and (\ref{interband}) show that if the dominating part of the 
$\langle L,I_2||\mathcal{M}||L,I_1\rangle$ matrix element is real then the inband electromagnetic transition should 
mainly be observed. 
On the contrary, if the imaginary part of the 
$\langle L,I_2||\mathcal{M}||L,I_1\rangle$ matrix element dominates then  
the interband electromagnetic transition should be in favor of the inband one.
These features are observed in the majority of partner bands in which lifetimes of the excited states have been measured.
The role of the specified handedness of the intrinsic states is then explained.
Staggering of the B(M1) value as a function of spin cannot be attributed only to the existence 
of the $|L\rangle$ and $|R\rangle$ states in a quantum system. 
This effect, as discussed in Ref.\cite{hamamoto}, relates to properties (here called structural composition) 
that can occur in addition with the chiral symmetry breaking. 
In the quoted paper, the Hamiltonian of the model consisting of a triaxially deformed core coupled to 
one proton particle and one neutron hole in the same single-$j$ shell was examined. 
Such a Hamiltonian, appropriate for the description of chiral bands built on two 
quasi-particle $\pi h_{11/2} \otimes \nu^{-1} h_{11/2}$ configuration, possesses additional invariances and 
corresponding quantum numbers. 
These quantum numbers have been used to derive the gamma selection rules which are based on the fact 
that the core angular momentum change for E2 and M1 transitions is $\Delta j_c \geq 2$. This special feature of core 
rotation is due to its triaxial deformation.  
Staggering of the B(M1) transition probabilities relates therefore to the structural composition of the laboratory states 
and can be explained by examining the $j_{\nu}$, $j_{\pi}$, $j_c$ quantum numbers in matrix elements 
$\langle I_2, \pm ,j_{\nu},j_{\pi},j_c||\mathcal{M}||I_1, \pm, j_{\nu},j_{\pi},j_c\rangle$. 
If some of the conditions given in Ref. \cite{hamamoto} were not fulfilled then B(M1) staggering should vanish. 
Nonetheless, other properties related to the existence of the $|L\rangle$ and $|R\rangle$ states 
could  still be observed.
Such a situation takes place in $^{102, 103, 104, 105, 106}$Rh nuclei where partner bands are expected to be built on 
the configuration consisting of different $j$-shell levels \cite{timar_06} \cite{vaman_04}, \cite{joshi}, as well as 
in the $^{135}$Nd isotope where partner 
bands are assumed to be built on the three (instead of two) quasiparticle configuration \cite{mukhopadhyay_07}.
Lifetimes of the excited states were measured in partner bands of $^{104}$Rh \cite {suzuki} and 
$^{135}$Nd \cite{mukhopadhyay_07} and, indeed, the staggering of B(M1) transition probability observed in those bands 
is very weak (compared to that for Cs isotopes) or absent. 
Moreover, in the case of $^{135}$Nd the corresponding transition probabilities in the partner bands are almost equal.
This points to the strong chiral symmetry breaking limit in $^{135}$Nd 
and reflects different structural composition than in Cs isotopes.
These examples indicate that B(M1) staggering should be a sensitive measure of triaxiality in two-quasiparticle 
single $j$-shell configurations. 
The dependence of the B(M1) staggering on triaxial deformation has been studied in terms of the CQPC model \cite{droste} 
where indeed the staggering effect vanishes rapidly when $\gamma$-deformation deviates from 30$^{\circ}$.
This fact supports the role of the structural composition in description of the electromagnetic properties of the partner bands.

\section{summary}
It follows from the above considerations that the chiral symmetry breaking in nuclear physics 
leads to the existence of two rotational bands, almost degenerated with similar 
electromagnetic properties. 
Staggering of the B(M1) value quoted in many papers is an additional property originating from the structural 
composition of a given nucleus.

This work was supported in part by the Polish Ministry of Science under Contract No.~N~N202~169736.

\end{document}